
\input harvmac
\input amssym.def
\input epsf
\noblackbox
\newcount\figno
\figno=0
\def\fig#1#2#3{
\par\begingroup\parindent=0pt\leftskip=1cm\rightskip=1cm\parindent=0pt
\baselineskip=11pt
\global\advance\figno by 1
\midinsert
\epsfxsize=#3
\centerline{\epsfbox{#2}}
\vskip 12pt
\centerline{{\bf Figure \the\figno} #1}\par
\endinsert\endgroup\par}
\def\figlabel#1{\xdef#1{\the\figno}}
\def\pano{\par\noindent}

\def\pmb#1{\setbox0=\hbox{#1}%
 \kern-.025em\copy0\kern-\wd0
 \kern.05em\copy0\kern-\wd0
 \kern-.025em\raise.0433em\box0 }
\font\cmss=cmss10
\font\cmsss=cmss10 at 7pt
\def\inbar{\,\vrule height1.5ex width.4pt depth0pt}
\def\half{{1\over 2}}
\def\rlx{\relax\leavevmode}
\def\Cop{\relax\,\hbox{$\inbar\kern-.3em{\rm C}$}}
\def\Rop{\relax{\rm I\kern-.18em R}}
\def\Nop{\relax{\rm I\kern-.18em N}}
\def\one{\relax{\rm 1\kern-.25em I}}
\def\Pop{\relax{\rm I\kern-.18em P}}
\def\Zop{\rlx\leavevmode\ifmmode\mathchoice{\hbox{\cmss Z\kern-.4em Z}}
 {\hbox{\cmss Z\kern-.4em Z}}{\lower.9pt\hbox{\cmsss Z\kern-.36em Z}}
 {\lower1.2pt\hbox{\cmsss Z\kern-.36em Z}}\else{\cmss Z\kern-.4em
 Z}\fi}


\def\N{{\cal N}}

\def\H{{\cal H}}

\def\A{{\cal A}}

\def\half{{1\over 2}}
\def\ie{{\it i.e.}}
\def\eg{{\it e.g.}}

\def\Vir{\rm Vir}

\def\Hbar{\bar{H}}

\def\suhat{\widehat{su}}
\def\sohat{\widehat{so}}

\def\Hbar{\bar{\H}}

\def\arrow{\rightarrow}


\def\sqr#1#2{{\vcenter{\vbox{\hrule height.#2pt
 \hbox{\vrule width.#2pt height#1pt \kern#1pt
 \vrule width.#2pt}\hrule height.#2pt}}}}

  
\def\makeblankbox#1#2{\hbox{\lower\dp0\vbox{\hidehrule{#1}{#2}%
   \kern -#1
   \hbox to \wd0{\hidevrule{#1}{#2}%
      \raise\ht0\vbox to #1{}
      \lower\dp0\vtop to #1{}
      \hfil\hidevrule{#2}{#1}}%
   \kern-#1\hidehrule{#2}{#1}}}%
}%
\def\hidehrule#1#2{\kern-#1\hrule height#1 depth#2 \kern-#2}%
\def\hidevrule#1#2{\kern-#1{\dimen0=#1\advance\dimen0 by #2\vrule  
    width\dimen0}\kern-#2}%
\def\openbox{\ht0=1.2mm \dp0=1.2mm \wd0=2.4mm  \raise 2.75pt  
\makeblankbox {.25pt} {.25pt}  }

\def\bun#1/#2{\leavevmode  
   \kern.1em \raise .5ex \hbox{\the\scriptfont0 #1}%
   \kern-.1em $/$%
   \kern-.15em \lower .25ex \hbox{\the\scriptfont0 #2}%
}

\def\opensquare{\ht0=3.4mm \dp0=3.4mm \wd0=6.8mm  \raise 2.7pt  
\makeblankbox {.25pt} {.25pt}  }  
  
  
\def\sector#1#2{\ {\scriptstyle #1}\hskip 1mm  
\mathop{\opensquare}\limits_{\lower 1mm\hbox{$\scriptstyle#2$}}\hskip
   1mm}    
  
\def\tsector#1#2{\ {\scriptstyle #1}\hskip 1mm  
\mathop{\opensquare}\limits_{\lower
   1mm\hbox{$\scriptstyle#2$}}^\sim\hskip  1mm}  





\lref\ABG{
C.~Angelantonj, R.~Blumenhagen, M.R.~Gaberdiel, {\it Asymmetric
orientifolds, brane supersymmetry breaking and non-BPS  branes},  
Nucl.\ Phys.\  {\bf B589}, 545 (2000); {\tt hep-th/0006033}.}

\lref\carlo{C.~Angelantonj, I.~Antoniadis, K.~Foerger, 
{\it Non-supersymmetric type I strings with zero vacuum energy},
Nucl.\ Phys.\ {\bf B555}, 116 (1999); {\tt hep-th/9904092}.}

\lref\bantayI{
P.~Bantay, {\it Characters and modular properties of permutation orbifolds},
Phys.\ Lett.\ {\bf B419}, 175 (1998); {\tt hep-th/9708120}.}

\lref\bantayII{
P.~Bantay, {\it Permutation orbifolds},
Nucl.\ Phys.\ {\bf B633}, 365 (2002); {\tt hep-th/9910079}.}

\lref\dongmason{K.~Barron, C.~Dong, G.~Mason, 
{\it Twisted sectors for tensor product vertex operator algebras
associated to permutation groups}, 
Commun.\ Math.\ Phys. {\bf 227}, 349 (2002); {\tt math.QA/9803118}.}

\lref\BFS{
L.~Birke, J.~Fuchs,  C.~Schweigert, {\it Symmetry breaking boundary
conditions and WZW orbifolds}, 
Adv.\ Theor.\ Math.\ Phys.\  {\bf 3}, 671 (1999); 
{\tt hep-th/9905038}.}

\lref\solit{J.~Fuchs, C.~Schweigert, {\it Solitonic sectors,
alpha-induction and symmetry breaking boundaries}, Phys.\ Lett.\
{\bf B490}, 163 (2000); {\tt hep-th/0006181}.} 

\lref\BRR{I.~Brunner, A.~Rajaraman, M.~Rozali, {\it D-branes on 
asymmetric orbifolds}, Nucl.\ Phys.\ {\bf B558}, 205 (1999); 
{\tt  hep-th/9905024}.} 

\lref\blumenhagen{R.~Blumenhagen, L.~G\"orlich, {\it Orientifolds of 
non-supersymmetric asymmetric orbifolds}, Nucl.\ Phys.\ {\bf B551},
601 (1999); {\tt hep-th/9812158}.} 

\lref\BGKL{
R.~Blumenhagen, L.~G\"orlich, B.~K\"ors, D.~L\"ust, {\it Asymmetric
orbifolds, noncommutative geometry and type I string vacua}, 
Nucl.\ Phys.\  {\bf B582}, 44 (2000); {\tt hep-th/0003024}.}

\lref\Cardy{
J.L.~Cardy, {\it Boundary conditions, fusion rules and the Verlinde
formula}, Nucl.\ Phys.\ {\bf B324}, 581 (1989).}

\lref\CGH{B.~Craps, M.R.~Gaberdiel, J.A.~Harvey, {\it Monstrous
branes}, to appear in Commun.\ Math.\ Phys., 
{\tt hep-th/0202074}.}

\lref\FSone{J.~Fuchs, C.~Schweigert, {\it Symmetry breaking
boundaries. I: General theory}, Nucl.\ Phys.\  {\bf B558}, 419 (1999);  
{\tt hep-th/9902132}.}

\lref\FStwo{J.~Fuchs, C.~Schweigert, {\it Symmetry breaking
boundaries. II: More structures, examples}, Nucl.\ Phys.\ {\bf B568}, 
543 (2000); {\tt hep-th/9908025}.}

\lref\QRS{T.~Quella, I.~Runkel, C.~Schweigert, {\it An algorithm for
twisted fusion rules}, {\tt math.qa/ 0203133}.} 

\lref\QS{T.~Quella, V.~Schomerus, {\it Symmetry breaking boundary
states and defect lines}, Journ.\ High Energy Phys.\ {\bf 0206}, 028
(2002); {\tt hep-th/0203161}.}

\lref\GaGa{M.R.~Gaberdiel, T.~Gannon, {\it Boundary states for
WZW models}, Nucl.\ Phys.\ {\bf B639}, 471 (2002); {\tt hep-th/0202067}.}

\lref\GRW{M.R.~Gaberdiel, A.~Recknagel, G.M.T.~Watts, {\it The
conformal boundary states for SU(2) at level 1}, Nucl.\ Phys.\ 
{\bf B626}, 344 (2002); {\tt hep-th/0108102}.}

\lref\Gannon{T.~Gannon, M.A.~Walton, {\it Heterotic modular
invariants and level-rank duality}, Nucl.\ Phys.\ {\bf B536}, 553
(1998); {\tt hep-th/9804040}.} 

\lref\PG{P.~Goddard, D.I.~Olive, {\it Kac-Moody and Virasoro algebras in
relation to quantum physics}, Int.\ J.\ Mod.\ Phys.\  {\bf A1}, 303
(1986).}

\lref\gutperle{M.~Gutperle, {\it Non-BPS D-branes and enhanced
symmetry in an asymmetric orbifold}, Journ.\ High Energy Phys.\ 
{\bf 0008}, 036 (2000); {\tt hep-th/0007126}.}

\lref\Harvey{J.A.~Harvey, {\it String duality and non-supersymmetric
strings}, Phys.\ Rev.\ {\bf D59}, 026002 (1999); 
{\tt hep-th/9807213}.} 

\lref\KKSone{
S.~Kachru, J.~Kumar, E.~Silverstein, {\it Vacuum energy
cancellation in a non-super\-sym\-met\-ric string}, 
Phys.\ Rev.\  {\bf D59}, 106004 (1999); {\tt hep-th/9807076}.}

\lref\KKStwo{  
S.~Kachru, E.~Silverstein, {\it 4d conformal theories and strings 
on orbifolds}, Phys.\ Rev.\ Lett.\  {\bf 80}, 4855 (1998); 
{\tt hep-th/9802183}.} 

\lref\klemmsch{
A.~Klemm, M.G.~Schmidt, {\it Orbifolds by cyclic permutations of
tensor product conformal field theories}, 
Phys.\ Lett.\  {\bf B245}, 53 (1990).}

\lref\kors{B.~K\"ors, {\it D-brane spectra of nonsupersymmetric,
asymmetric orbifolds and  nonperturbative contributions to the
cosmological constant}, Journ.\ High Energy Phys.\ {\bf 9911}, 028
(1999); {\tt hep-th/9907007}.}

\lref\kreuschell{M.~Kreuzer, A.N.~Schellekens,
{\it Simple currents versus orbifolds with discrete torsion: a
complete classification}, 
Nucl.\ Phys.\  {\bf B411}, 97 (1994); {\tt hep-th/9306145}.}

\lref\MMSI{       
J.~M.~Maldacena, G.~W.~Moore, N.~Seiberg,
{\it Geometrical interpretation of D-branes in gauged WZW models}, 
Journ.\ High Energy Phys.\ {\bf 0107}, 046 (2001); {\tt hep-th/0105038}. 
}

\lref\maloney{
A.~Maloney, E.~Silverstein, A.~Strominger, {\it De Sitter space in
noncritical string theory}, {\tt hep-th/0205316}.}

\lref\NSV{K.S.~Narain, M.H.~Sarmadi, C.~Vafa, {\it Asymmetric
orbifolds}, Nucl.\ Phys.\ {\bf B288}, 551 (1987).}

\lref\RS{A.~Recknagel, V.~Schomerus, {\it Boundary deformation theory 
and moduli spaces of D-branes}, Nucl.\ Phys.\ {\bf B545}, 233 (1999);  
{\tt hep-th/9811237}.}

\lref\Silver{E. Silverstein, {\it (A)dS backgrounds from asymmetric
orientifolds}, {\tt hep-th/0106209}.}

\lref\SYone{A.N.~Schellekens, S.~Yankielowicz, {\it Extended 
chiral algebras and modular invariant partition functions}, Nucl.\ 
Phys.\ {\bf B327}, 673 (1989).}           

\lref\intril{K. Intriligator, {\it Bonus symmetry in conformal
field theory}, Nucl.\ Phys.\ {\bf B332}, 541 (1990).}

\lref\SYtwo{A.N.~Schellekens, S.~Yankielowicz, {\it Simple 
currents, modular invariants and fixed points}, Int.\ J.\ Mod.\ Phys.\ 
{\bf A5}, 2903 (1990).} 

\lref\tseng{L.-S.~Tseng, {\it A note on c=1 Virasoro boundary states 
and asymmetric shift orbifolds},  Journ.\ High Energy Phys.\ 
{\bf 0204}, 051 (2002); {\tt hep-th/0201254}.}

\lref\vafawitten{C.~Vafa, E.~Witten, {\it Dual string pairs with N=1
and N=2 supersymmetry in four dimensions},
Nucl.\ Phys.\ Proc.\ Suppl.\ {\bf 46}, 225 (1996); 
{\tt hep-th/9507050}.}

\lref\recknagel{A.~Recknagel, {\it Permutation branes}, 
{\tt hep-th/0208119}.}

\lref\OOY{H.~Ooguri, Y.~Oz, Z.~Yin, {\it D-Branes on Calabi-Yau spaces
and their mirrors}, Nucl.\ Phys.\ {\bf B477}, 407 (1996);
{\tt hep-th/9606112}.}


\Title{\vbox{
\hbox{hep-th/0210137}
\hbox{KCL-MTH-02-23}
\hbox{DAMTP-2002-102}}}
{\vbox{\centerline{D-branes in an asymmetric orbifold}}}
\centerline{Matthias R.\ Gaberdiel\footnote{$^\ddagger$}{On leave of
absence from: Department of Mathematics, King's College London,
Strand, London WC2R 2LS, UK, {\tt e-mail: mrg@sns.ias.edu,
mrg@mth.kcl.ac.uk}}$^{,a}$ and  
Sakura Sch\"afer-Nameki\footnote{$^\star$}{{\tt e-mail: 
S.Schafer-Nameki@damtp.cam.ac.uk}}$^{,b}$}
\bigskip
\centerline{\it $^a$School of Natural Sciences, Institute for Advanced
Study}
\centerline{\it Princeton, NJ 08540, USA}
\medskip
\centerline{\it $^b$Department of Applied Mathematics and Theoretical
Physics}  
\centerline{\it Wilberforce Road, Cambridge CB3 OWA, U.K.}
\smallskip
\vskip2cm
\centerline{\bf Abstract}
\bigskip
\noindent
We consider the asymmetric orbifold that is obtained by acting with
T-duality on a 4-torus, together with a shift along an extra circle. 
The chiral algebra of the resulting theory has non-trivial outer
automorphisms that act as permutations on its simple factors. These
automorphisms play a crucial role for constructing D-branes 
that couple to the twisted sector of the orbifold.
\bigskip

\Date{10/2002}


\newsec{Introduction}

Asymmetric orbifolds \refs{\NSV} have played a prominent
r\^ole in constructing string backgrounds with phenomenologically
interesting features 
\refs{\KKSone,\KKStwo,\Harvey,\Silver,\maloney,\blumenhagen,\carlo,
\ABG,\kors,\BGKL}. In particular, they provide interesting
non-supersymmetric string theories as well as dS-like backgrounds.  
From a conformal field theoretic point of view, asymmetric orbifolds
are of interest since they provide a way to construct new non-diagonal,
so-called heterotic modular invariants, see \eg\ \refs{\Gannon} and
references therein. 

Despite the large literature on asymmetric orbifolds, little
is known about D-branes in these backgrounds. Some results for
specific orbifolds have been obtained in
\refs{\BRR,\gutperle,\tseng,\CGH}, but a complete understanding still
seems to be lacking. One natural candidate for a symmetry that acts
differently on left and right moving degrees of freedom is
T-duality. Orbifolds including an element of the T-duality group have
been considered in \refs{\KKSone,\KKStwo, \Harvey,\blumenhagen,\carlo}
in the context of non-supersymmetric string compactifications. In this
paper we will construct the D-brane boundary states in a bosonic
relative of   these models. While this model is only a toy model, some
of the techniques we shall discuss will generalise to orbifolds of the 
superstring. 

Our construction also possesses a few novel conformal field theoretic 
features. Although it is rather straightforward to construct a class
of boundary states that are invariant under the orbifold action, it is
typically difficult to construct D-branes that couple to states in the
twisted sector of an asymmetric orbifold. In order to obtain such
boundary states in our example, we will have to consider D-branes that
preserve the maximal chiral symmetry only up to an automorphism of the
symmetry algebra. Symmetry breaking boundary conditions, in particular
for simple chiral algebras, have been studied {\it e.g.}\ in 
\refs{\OOY,\BFS,\FSone,\FStwo,\solit,\GaGa,\MMSI,\QS}. The orbifold chiral
algebra that we shall encounter here consists of powers of a simple
chiral algebra, and thus allows for extra outer automorphisms, which
act as permutations on the factors (compare
\refs{\recknagel})\footnote{$^\dagger$}{Orbifolds of a closed string
theory by a permutation group have been discussed \eg\ in  
\refs{\klemmsch,\bantayI,\bantayII,\dongmason}.}. 
In fact we will be able to construct all boundary states that preserve
the orbifold chiral algebra up to an arbitrary automorphism. Taken
together, these boundary states couple to all different sectors of the
theory. 
\smallskip

The plan of the paper is as follows. In section~2 we describe the
model we are discussing and fix the notation. In section~3 we explain
that the asymmetric orbifold can alternatively be described as a
simple current extension of the $\suhat(2)_1^5$ diagonal theory. The
permutation branes of this diagonal theory are constructed, and the
NIM-rep property of the twisted fusion algebra is checked.  In
section~4 we then construct permutation branes for the asymmetric
orbifold and verify Cardy's condition. Section~5 explains
how to generalise the construction to `conformal' boundary states, and  
how superpositions of conventional Dirichlet- and  
Neumann-branes fit into our picture. Our conclusions are contained in
section~6. There are three appendices where some more technical 
details are spelled out.

\newsec{The setup}

In this paper we shall be interested in the following asymmetric
orbifold. Consider the toroidal compactification of the (bosonic)
string on a 4-torus at the SO(8) point. At this point in moduli space,
the theory is simply the (diagonal) level $k=1$ $\sohat(8)_1$
Kac-Moody theory. Let us choose the Cartan-Weyl basis for the
generators of $\sohat(8)_1$, and let us denote the `Cartan generators'
by $H^i_n$, while the `root generators' are $E^\alpha_n$. We are 
interested in the orbifold action  that acts on the left-moving
currents as  
\eqn\Tonsoeight{
g_L\,:\, \ E_n^{\pm \alpha}\, \arrow\, - E_n^{\mp\alpha}\,,\quad
H^i_n\,\arrow\, -H^i_n\,, }
while the action on the right-movers is trivial, $g_R=\one$. The
invariant algebra for the left-movers is then  
\eqn\Tinvariantalgebra{
\left<\left\{\, E^{\alpha} -  E^{-\alpha}\, | \,
\alpha\in\Delta^+ \,\right\}\, \right>\,,}
which is isomorphic to $\A_L^g\,=\, \sohat(4)_1\oplus \sohat(4)_1$. The
invariant algebra for the right-movers is obviously 
$\A_R^g\,=\, \sohat(8)_1$. The action \Tonsoeight\ can be thought of
as some form of T-duality along the four torus 
directions\footnote{$^\ddagger$}{As is explained in \refs{\CGH}, this
transformation differs from the usual T-duality transformation by a
rotation on the right-movers.}.

In order to describe the action of $g_L$ on the four different
representations of $\sohat (8)_1$ it is useful to decompose them with
respect to $\A_L^g$. Let us label the representations of the
$\sohat(4n)_1$ chiral algebra by $o,v,s,c$, and let us distinguish the
two $so(4)$ copies by $I$ and $II$. Then the relevant decomposition is 
\eqn\soeightintosofour{
\eqalign{
& \ \quad\,  \psi=1\quad\quad\qquad\,   \psi= -1\cr
\H_{o}^{\sohat (8)}\,=& \, \left(\H^I_o\otimes \H^{II}_o\right)
\,\oplus\, \left(\H^I_v\otimes \H^{II}_v\right) \cr  
\H_{v}^{\sohat (8)}\,=& \, \left(\H^I_o\otimes \H^{II}_v\right)
\,\oplus\, \left(\H^I_v\otimes \H^{II}_o\right) \cr  
\H_{s}^{\sohat (8)}\,=& \, \left(\H^I_s\otimes \H^{II}_s\right)
\,\oplus\, \left(\H^I_c\otimes \H^{II}_c\right) \cr  
\H_{c}^{\sohat (8)}\,=& \, \left(\H^I_s\otimes \H^{II}_c\right)
\,\oplus\, \left(\H^I_c\otimes \H^{II}_s\right)\,.
}}
On these representations $g_L$ then acts as $\psi=\pm 1$.

The orbifold defined by $g$ does not satisfy the level matching
condition. Indeed, since we invert four left-moving modes without
modifying the right-movers, the energy shifts of the left- and  
right-moving ground state in the twisted sector are
\eqn\levelmismatch{
\Delta_L\,=\, {4\over 48}+ {4\over 24}\,=\, {1\over 4}\,,\qquad
\Delta_R\,=\, 0\,,
}
and thus the action by $g$ is not consistent by itself. The apparent
level-matching problem \levelmismatch\ of the orbifold by $g$ can be
cured by introducing additional shifts in the Narain lattice. Let us 
denote the inequivalent choices as in \vafawitten\ by 
\eqn\Ashifts{
A_1= \left({1\over 2\sqrt{2}}, {1\over 2\sqrt{2}} \right)\,,\quad
A_2= \left({1\over \sqrt{2}}, 0\right)\,, \quad 
A_3= \left({1\over 2\sqrt{2}}, -{1\over 2\sqrt{2}} \right)\,.}
On states with momentum and winding 
$(p_L,p_R)=({n+w\over\sqrt{2}},{n-w\over\sqrt{2}})$ these shifts act
via the phase $\exp(2\pi i p\cdot A)$, \ie\ as 
\eqn\shiftaction{
\left(-1\right)^{w}\,, \quad (-1)^{w+n}\,, \quad
(-1)^{n}\,,}
respectively.

For the supersymmetric version of this theory, a resolution of the 
level-mismatch has been discussed earlier in the literature
\refs{\KKSone,\KKStwo,\blumenhagen,\carlo}. To this end, one combines the  
orbifold action with a shift by $A_1$ along all four torus
directions. This resolves the level-matching problem since the shift
changes the right-moving ground state energy in the twisted sector by  
$\Delta_R= 4 {A_{1,R}^2\over 2}\,=\, {1\over 4}$, while it does not
contribute to the ground state energy of the left-movers, since the
left-movers are reflected (and the shift can therefore be undone).
If this resolution is applied to the bosonic theory in question, the
resulting theory becomes actually left-right symmetric, and is
therefore not of immediate interest. 
\bigskip

In order to obtain a genuinely asymmetric orbifold, we therefore
choose another way of satisfying the level matching condition which is
similar to what was proposed in \Harvey. Let us compactify the theory
on an additional circle for which we choose the radius to take the
self-dual value, and combine $g$ with an $A_2$ shift along this
additional circle,  
\eqn\cftf{
\left. T^4_{so(8)}\, \times\, T^1_{su(2)}\,\right/ f=\left(\,
(-1)^4_L\,, \ A_2\, \right)\,.
}
The additional shift changes the left-moving ground state energy by  
\eqn\deltal{
\Delta_{L}\,=\, {(A_2)^2\over 2}\,=\, {1\over 4}\,,\qquad
\Delta_{R}\,=\, 0\,,}
and therefore resolves \levelmismatch. The shift $A_2$ of \Ashifts\
has the action \vskip1pt
\eqn\sutwoinuone{
\eqalign{
& \quad\quad  \psi'=1\qquad \qquad\quad\  \psi'=-1\cr
\H_{T^1_{su(2)}}\,=& \quad \H_{0,1}
\otimes \Hbar_{0,1} \quad \oplus\quad \H_{1,1}\otimes \Hbar_{1,1} \,,
}}
where $\H_{j,1}$ for $j=0,1$ denote the two irreducible highest weight
representations of $\suhat(2)_1$. The spectrum of the resulting theory
is then genuinely asymmetric, and is explicitly given as 
\eqn\HUHT{\eqalign{
\H_U\,=\,   \quad & \left(\, 
\left(\, \H^I_o\otimes \H^{II}_o\, \right) \,\otimes \, \Hbar^{so(8)}_o   
\,\oplus \, \left(\, \H^I_o\otimes \H^{II}_v\, \right) \,\otimes \,
\Hbar^{so(8)}_v\right.\cr    
& \oplus \, \left.\left(\, \H^I_s\otimes \H^{II}_s\, \right)
\,\otimes \, \Hbar^{so(8)}_s    
\, \oplus \, \left(\, \H^I_s\otimes \H^{II}_c\, \right) \,\otimes \,
\Hbar^{so(8)}_c    
\,\right)\, \otimes \H_{0,1}\otimes \Hbar_{0,1} \cr
 \quad \oplus & \left(\, 
\left(\, \H^I_v\otimes \H^{II}_v\, \right) \,\otimes \,
\Hbar^{so(8)}_o    
\, \oplus \, \left(\, \H^I_v\otimes \H^{II}_o\, \right) \,\otimes \,
\Hbar^{so(8)}_v \right.\cr   
& \oplus \,  \left.\left(\, \H^I_c\otimes \H^{II}_c\, \right)
 \,\otimes \, \Hbar^{so(8)}_s    
\, \oplus \, \left(\, \H^I_c\otimes \H^{II}_s\, \right) \,\otimes \,
 \Hbar^{so(8)}_c    
\,\right)\, \otimes \H_{1,1} \otimes \Hbar_{1,1} \cr
\H_T\,=\,   \quad & \left(\, 
(\H^I_c\otimes \H^{II}_v)\, \otimes \, \Hbar^{so(8)}_v \,
\oplus\, (\H^I_c\otimes \H^{II}_o)\, \otimes \, \Hbar^{so(8)}_o
\right. \cr
& \oplus \, \left. 
(\H^I_v\otimes \H^{II}_c)\, \otimes \, \Hbar^{so(8)}_c \,
\oplus\, (\H^I_v\otimes \H^{II}_s)\, \otimes \, \Hbar^{so(8)}_s
\right)\, \otimes \, (\H_{0,1}\otimes \Hbar_{1,1}) \cr 
\quad \oplus & \ \, \left(\,
(\H^I_o\otimes \H^{II}_s)\, \otimes \, \Hbar^{so(8)}_c \,
\oplus\, (\H^I_o\otimes \H^{II}_c)\, \otimes \, \Hbar^{so(8)}_s
\right. \cr
& \oplus \, \left. 
(\H^I_s\otimes \H^{II}_o)\, \otimes \, \Hbar^{so(8)}_v \,
\oplus\, (\H^I_s\otimes \H^{II}_v)\, \otimes \, \Hbar^{so(8)}_o
\right)\, \otimes \, (\H_{1,1}\otimes \Hbar_{0,1}) 
\,,}}
where $\H_U$ and $\H_T$ denote the untwisted and twisted sectors,
respectively. 
Here the representations have been written in terms of the chiral
algebras 
\eqn\orbifoldchiral{
\A_L\,=\, \sohat(4)_1\oplus \sohat(4)_1\oplus \suhat(2)_1\,,\qquad 
\A_R\,=\, \sohat(8)_1\oplus \suhat(2)_1\,.}
The maximal diagonal chiral subalgebra is therefore 
\eqn\diagonal{
\A_{diag}\,=\,  \sohat(4)_1\oplus \sohat(4)_1\oplus \suhat(2)_1
\cong \suhat(2)_1^{\oplus 5} \,.}
For definiteness, we identify the first two $su(2)$s with the first
$so(4)$ in \diagonal, and the second two $su(2)$s with the second 
$so(4)$. Furthermore, we identify the representations of 
$\sohat(4)_1$ with those of $\suhat(2)_1\oplus \suhat(2)_1$ as 
\eqn\sofoursutwo{
O \simeq (0,0)\,, \qquad
V \simeq (1,1)\,, \qquad
S \simeq (1,0)\,, \qquad
C \simeq (0,1)\,.}

\newsec{Permutation-twisted boundary states}

The main aim of this paper is to construct all the D-branes of the
asymmetric orbifold that preserve \diagonal\ up to automorphisms. This
is to say, we want to construct the boundary states that preserve
\eqn\generalglue{
\left(\, J_{n}^a - \sigma( \bar{J}_{-n}^a)\, \right)\, 
|\!|\,{\bf \alpha}\,\rangle\!\rangle\,=\, 0\,,}
where $J^a_n$ is a current in \diagonal, and $\sigma$ is an arbitrary
automorphism of $\A_{diag}$. In particular, we are interested in
D-branes that couple to the twisted sector states of the asymmetric
orbifold. It is obvious from \HUHT\ that the boundary states that
involve Ishibashi states from the twisted sector components can only
arise for automorphisms $\sigma$ that are outer. The group of outer
automorphisms of $\A_{diag}$ is isomorphic to the permutation group in
five objects, $S_5$, where $\sigma\in S_5$ acts on the $\suhat(2)_1$
factors as  
\eqn\sigmaaction{
\sigma\,:\quad  \suhat(2)_1^{(i)}\, \arrow\,  
\suhat(2)_1^{(\sigma i)}\,. }
For fixed $\sigma\in S_5$, the boundary states that satisfy
\generalglue\ can be written in terms of the $\sigma$-twisted
Ishibashi states, 
\eqn\boundexp{
|\!| {\bf \alpha} \rangle\!\rangle^\sigma = 
\sum_{\bf l} B_{{\bf \alpha},{\bf l}} \,
| {\bf l}\rangle\!\rangle^\sigma \,,
}
where ${\bf \alpha}$ labels the different $\sigma$-twisted boundary
states, while $| {\bf l}\rangle\!\rangle^\sigma$ denotes the 
$\sigma$-twisted Ishibashi state that originates from the
representation of $\A_{diag}\otimes\A_{diag}$ 
\eqn\twistedweights{
(l_1,l_2,l_3,l_4,l_5)\otimes \sigma^{-1}\,(l_1,l_2,l_3,l_4,l_5)\,,
\quad l_i=0,1\,.}

In principle one could now go ahead and determine the set of
$\sigma$-twisted Ishibashi states for each element $\sigma\in S_5$,
and then make an ansatz for the boundary states that can be written as
a linear superposition of these Ishibashi states. However, the
discussion simplifies drastically if we relabel our theory as
follows. Let us redefine the right-moving chiral algebra $\A^R_{diag}$
by exchanging the second and fifth $\suhat(2)_1$ factors. (In terms of
the boundary states in \generalglue\ this amounts to replacing
$\sigma$ by $\sigma (25)$.) After a little calculation one then
finds that, with respect to this redefined chiral algebra, the
spectrum of the theory simply becomes
\eqn\Jdecomp{
{\cal H}= \bigoplus_{J_{\bf l}=+1} {\bf l} \otimes {\bf l}\quad
    \oplus\quad 
    \bigoplus_{J_{\bf l}=+1} {\bf l} \otimes J {\bf l} \,. 
}
Here ${\bf l}=(l_1,l_2,l_3,l_4,l_5)$ labels representations of
$\suhat(2)_1^5$, and the sum runs over all such representations for
which the eigenvalue $J_{\bf l}$, defined by 
\eqn\Jeigen{
J_{\bf l} = (-1)^{l_1+l_3+l_4+l_5}\,,
}
equals $+1$. Furthermore, the `simple current' $J$ acts on ${\bf l}$
as  
\eqn\Jaction{
J (l_1,l_2,l_3,l_4,l_5) = (l_1+1,l_2,l_3+1,l_4+1,l_5+1)\,,
}
where addition is understood modulo $2$. The first sum in \Jdecomp\
is the subsector of the diagonal $\suhat(2)_1^5$ theory that has
eigenvalue $+1$ under the action of $J$, while the second sum
describes the $J$-twisted sector. We can therefore think of this
theory as a simple current extension of the original diagonal
$\suhat(2)_1^5$ theory (see for example
\refs{\SYone,\intril,\SYtwo,\kreuschell}). As an aside it may be worth
pointing out that the first and second sum of \Jdecomp\ (\ie\ the
untwisted and  twisted sector of the simple current extension) do not
correspond to the untwisted and twisted sector of the original
orbifold in \HUHT, respectively. In order to construct D-branes for
the theory in question, it is now convenient to begin with
constructing permutation-twisted branes for the diagonal
$\suhat(2)_1^5$ theory.

\subsec{Permutation-twisted boundary states for $su(2)_1^n$}

The construction of the permutation-twisted boundary states can
actually be formulated more generally for the diagonal $\suhat(2)_1^n$ 
theory, and we shall therefore be more general in the following
(compare also \recknagel). In particular, we shall show that the
boundary states we construct satisfy the various Cardy conditions
\refs{\Cardy}.  

Suppose $\sigma$ is a permutation in $S_n$, and ${\bf \hat{m}}$ is a 
$\sigma$-twisted representation of $\suhat(2)_1^n$. For each such  
${\bf \hat{m}}$ we define the $\sigma$-twisted boundary state as 
\eqn\sigmaCardy{
|\!| {\bf \hat{m}} \rangle\!\rangle^\sigma = 
\sum_{\bf l} {\hat{S}^\sigma_{{\bf \hat{m}},{\bf l}} \over 
      \sqrt{S_{{\bf 0},{\bf l}}}} \, 
                   | {\bf l}  \rangle\!\rangle^\sigma \,,
}
where the sum runs over the sectors ${\bf l}=(l_1,\ldots,l_n)$ of
the diagonal $\suhat(2)_1^n$ theory for which 
$\sigma({\bf l})={\bf l}$, $S_{{\bf 0},{\bf l}}$ is a product of $n$
$\suhat(2)_1$ $S$-matrices, and 
$\hat{S}^\sigma_{{\bf \hat{m}},{\bf l}}$ is the $\sigma$-twisted
$S$-matrix. In order to define the latter, consider the `twining
character' 
\eqn\twining{
\chi_{\bf l}^{(\sigma)}(\tau) = 
\Tr_{\H_{\bf l}} (\sigma \, q^{L_0-c/24})\,, \qquad 
q=e^{2\pi i \tau}\,, 
}
which is non-zero if and only if ${\bf l}$ is invariant under $\sigma$, 
$\sigma({\bf l})={\bf l}$. Under the $S$-modular transformation these
twining characters transform into characters of $\sigma$-twisted
representations\footnote{$^\star$}{A twisted representation of a
chiral algebra is, by definition, the same as an untwisted
representation of the corresponding twisted algebra, see \refs{\PG}
for an introduction.},
\eqn\twinmod{
\chi_{\bf l}^{(\sigma)}(-1/\tau) = \sum_{{\bf \hat{m}}}
\hat{S}^\sigma_{{\bf \hat{m}},{\bf l}} \,
\chi_{\bf \hat{m}}(\tau)\,,}
where 
\eqn\twischar{
\chi_{\bf \hat{m}}(\tau) = \Tr_{\H_{\bf \hat{m}}} (q^{L_0-c/24})\,,
\qquad  q=e^{2\pi i \tau}\,.}
The number of $\sigma$-twisted representations always equals the
number of $\sigma$-invariant representations, and, in fact, 
$\hat{S}^\sigma_{{\bf \hat{m}},{\bf l}}$ is actually a unitary matrix. 

The formula \sigmaCardy\ is a natural generalisation of the formula
proposed in \refs{\GaGa,\BFS}. In the following we want to show that
it satisfies the Cardy condition. The Cardy condition states that the
overlap of two boundary states must give rise, after an $S$-modular
transformation, to a positive integer linear combination of (twisted)
characters of the chiral algebra. The calculation of the overlap
between two boundary states can be reduced to the overlap of two
Ishibashi states. The overlap between the latter can be easily
calculated, and it is given as 
\eqn\sigmatauoverlap{
{}^\sigma\langle\!\langle \,{\bf l}\,|
q^{{1\over 2}(L_0+\bar L_0 - c/12)} |\,{\bf k}\,\rangle\!\rangle^\tau
=\, \delta_{{\bf l},{\bf k}}\, \chi^{(\sigma\tau^{-1})}_{\bf l}(q)
\,,}
where $\chi^{(\sigma\tau^{-1})}_{\bf l}(q)$ is the twining character
\twining. Since only the $\sigma$-invariant states contribute to the
twining character, it is obvious that 
\eqn\twiningident{
\chi^{(\sigma)}_{\bf l}(q) = \chi^{(\sigma^{-1})}_{\bf l}(q)\,.}
Furthermore, since each permutation commutes with $L_0$, the identity
\eqn\twiningidentp{
\chi^{(\sigma \tau)}_{\bf l}(q) = \chi^{(\tau\sigma)}_{\bf l}(q)}
holds. In particular, it therefore follows that \sigmatauoverlap\ also 
equals  
\eqn\sigmatauoverlapp{
{}^\sigma\langle\!\langle \,{\bf l}\,|
q^{{1\over 2}(L_0+\bar L_0 - c/12)} |\,{\bf k}\,\rangle\!\rangle^\tau
=\, \delta_{{\bf l},{\bf k}}\, \chi^{(\sigma\tau^{-1})}_{\bf l}(q)
=\, \delta_{{\bf l},{\bf k}}\, \chi^{(\tau\sigma^{-1})}_{\bf l}(q)
=\, \delta_{{\bf l},{\bf k}}\, \chi^{(\sigma^{-1}\tau)}_{\bf l}(q)\,.}
With these preparations it is now immediate to write down the overlap
between two boundary states,
\eqn\NIMsutwo{
{}^\sigma \langle\!\langle {\bf \hat{n}} |\!| 
q^{{1\over 2}(L_0+\bar L_0 - c/12)} |\!|
{\bf \hat{m}} \rangle\!\rangle^\tau  = 
\sum_{{\bf \hat{p}}} \sum_{\bf l}
{\left(\hat{S}^\sigma_{{\bf \hat{n}},{\bf l}}\right)^\ast 
\hat{S}^{\sigma \tau^{-1}}_{{\bf \hat{p}},{\bf l}} 
\hat{S}^\tau_{{\bf \hat{m}},{\bf l}}
\over
S_{{\bf 0},{\bf l}}} \, \chi_{{\bf \hat{p}}}(\tilde{q}) 
= \sum_{{\bf \hat{p}}} 
{}^{(\sigma,\tau)}{\cal N}_{{\bf \hat{p}},{\bf \hat{m}}}^{{\bf \hat{n}}}
\, \chi_{{\bf \hat{p}}}(\tilde{q}) \,,}
where ${\bf \hat{p}}$ labels the set of $\sigma \tau^{-1}$-twisted 
representations of $\suhat(2)_1^n$, and 
$\tilde{q}=e^{-2\pi i /\tau}$. The Cardy condition is thus satisfied 
provided that 
\eqn\twistedfusion{
{}^{(\sigma,\tau)}{\cal N}_{{\bf \hat{p}},{\bf \hat{m}}}^
{{\bf \hat{n}}} =  \sum_{\bf l}
{\left(\hat{S}^\sigma_{{\bf \hat{n}},{\bf l}} \right)^\ast 
\hat{S}^{\sigma \tau^{-1}}_{{\bf \hat{p}},{\bf l}} 
\hat{S}^\tau_{{\bf \hat{m}},{\bf l}} 
\over
S_{{\bf 0},{\bf l}}} \,}
are non-negative integers; this can easily be confirmed explicitly
case by case. It is clear from the results of appendix~B that our 
formula for the boundary states agrees with the more explicit formula
given in \refs{\recknagel}; the fact that our boundary states satisfy
the Cardy condition follows then also from the analysis given
there. Finally, we have given a more abstract proof of this property
in appendix~B. 

It is very tempting to identify \twistedfusion\ with the fusion rules
describing the fusion of the $\tau$-twisted representation 
${\bf \hat{m}}$ with the  $\sigma\tau^{-1}$-twisted 
representation ${\bf \hat{p}}$ to give the  $\sigma$-twisted
representation ${\bf \hat{n}}$. Formula \twistedfusion\ generalises
then the Verlinde formula to twisted fusion rules; it is a natural
further generalisation of the formula proposed in \refs{\GaGa} (see
also \refs{\solit,\QRS}). For a specific example we have checked that
\twistedfusion\ does indeed describe the twisted fusion rules; this is
described in appendix~C.

\subsec{The NIM-rep property}

If the integers \twistedfusion\ describe the twisted fusion
rules, they must define a non-negative integer matrix 
representation of the fusion rules (or NIM-rep for short). This is to
say, the matrices \twistedfusion\ must satisfy 
\eqn\NIMrepprop{
\sum_{\bf \hat{m}}\, 
{}^{(\sigma,\tau)}{\cal N}^{\bf\hat{n}}_{{\bf\hat{p}},{\bf \hat{m}}}\ 
{}^{(\tau,\rho)}{\cal N}^{\bf\hat{m}}_{{\bf\hat{q}},{\bf \hat{k}}}\,
=\,   
\sum_{\bf \hat{r}}\, 
{}^{(\sigma,\rho)}{\cal N}^{\bf\hat{n}}_{{\bf\hat{r}},{\bf \hat{k}}} \
{}^{(\sigma\rho^{-1}, 
\tau\rho^{-1})}{\cal N}^{\bf\hat{r}}_{{\bf\hat{p}},{\bf \hat{q}}}\,. 
}
Here ${\bf \hat{k}}$ and ${\bf \hat{n}}$  are $\rho$-twisted and
$\sigma$-twisted representations, respectively, and the sum on the
left hand side runs over all $\tau$-twisted representations 
${\bf \hat{m}}$, while the sum on the right hand side runs over all 
$\sigma\rho^{-1}$-twisted representations ${\bf \hat{r}}$. 
Furthermore, ${\bf \hat{p}}$ and ${\bf \hat{q}}$  are
$\sigma\tau^{-1}$-twisted and $\tau\rho^{-1}$-twisted representations, 
respectively. 

To prove \NIMrepprop\ we write the left hand side as 
\eqn\NIMproof{\eqalign{
\sum_{\bf \hat{m}}\, 
{}^{(\sigma,\tau)}{\cal N}^{\bf\hat{n}}_{{\bf\hat{p}},{\bf \hat{m}}}\ 
{}^{(\tau,\rho)}{\cal N}^{\bf\hat{m}}_{{\bf\hat{q}},{\bf \hat{k}}}\,
&=\, \sum_{\bf \hat{m}}\, \sum_{{\bf l},{\bf l}'} 
{\left(\hat{S}^\sigma_{{\bf \hat{n}},{\bf l}}\right)^\ast 
 \hat{S}^\tau_{{\bf \hat{m}},{\bf l}} 
 \hat{S}^{\sigma \tau^{-1}}_{{\bf \hat{p}},{\bf l}} \over
  S_{{\bf 0},{\bf l}}}
{\left(\hat{S}^\tau_{{\bf \hat{m}},{\bf l}'} \right)^\ast
 \hat{S}^\rho_{{\bf \hat{k}},{\bf l}'} 
 \hat{S}^{\tau\rho^{-1}}_{{\bf \hat{q}},{\bf l}'} \over
  S_{{\bf 0},{\bf l}'}}
\cr
&=\, \sum_{{\bf l}} 
{\left(\hat{S}^\sigma_{{\bf \hat{n}},{\bf l}}  \right)^\ast
 \hat{S}^{\sigma \tau^{-1}}_{{\bf \hat{p}},{\bf l}} \over
  S_{{\bf 0},{\bf l}}}
{ \hat{S}^\rho_{{\bf \hat{k}},{\bf l}} 
 \hat{S}^{\tau\rho^{-1}}_{{\bf \hat{q}},{\bf l}} \over
  S_{{\bf 0},{\bf l}}}\,,
}}
where the sum over ${\bf l}$ in the first line extends over all
$\sigma$- and $\tau$-invariant representations, while the sum over
${\bf l}'$ runs over all $\tau$ and $\rho$-invariant representations. 
In going to the second line we have used that $\hat{S}^\tau$ is
unitary; the sum over ${\bf l}$ in the second line extends over all
representations that are simultaneously $\sigma$-,$\tau$- and
$\rho$-invariant. \NIMproof\ has to equal the right hand side of
\NIMrepprop 
\eqn\NIMproofcont{\eqalign{
\sum_{\bf \hat{r}}\, 
{}^{(\sigma,\rho)}{\cal N}^{\bf\hat{n}}_{{\bf\hat{r}},{\bf \hat{k}}}\ 
{}^{(\sigma\rho^{-1},
\tau\rho^{-1})}{\cal N}^{\bf\hat{r}}_{{\bf\hat{p}},{\bf \hat{q}}}\,
&=\, \sum_{\bf \hat{r}}\, \sum_{{\bf l},{\bf l}'} 
{\left(\hat{S}^\sigma_{{\bf \hat{n}},{\bf l}}\right)^\ast
 \hat{S}^\rho_{{\bf \hat{k}},{\bf l}} 
 \hat{S}^{\sigma \rho^{-1}}_{{\bf \hat{r}},{\bf l}} \over
  S_{{\bf 0},{\bf l}}}
{\left(\hat{S}^{\sigma\rho^{-1}}_{{\bf \hat{r}},{\bf l}'} \right)^\ast
 \hat{S}^{\sigma\tau^{-1}}_{{\bf \hat{p}},{\bf l}'} 
 \hat{S}^{\tau\rho^{-1}}_{{\bf \hat{q}},{\bf l}'} \over
  S_{{\bf 0},{\bf l}'}}
\cr
&=\, \sum_{\bf l} 
{\left(\hat{S}^\sigma_{{\bf \hat{n}},{\bf l}}\right)^\ast 
 \hat{S}^\rho_{{\bf \hat{k}},{\bf l}}\over 
  S_{{\bf 0},{\bf l}}}
{ \hat{S}^{\sigma\tau^{-1}}_{{\bf \hat{p}},{\bf l}} 
 \hat{S}^{\tau\rho^{-1}}_{{\bf \hat{q}},{\bf l}} \over
  S_{{\bf 0},{\bf l}}}
\,,
}}
where the sum in the last line is over $\sigma$-, $\tau$- and
$\sigma\rho^{-1}$-invariant, and hence also $\rho$-invariant,
representations ${\bf l}$. This then agrees with \NIMproof.

\newsec{Permutation-twisted D-branes in the asymmetric orbifold}

Let us now return to the description of the D-branes in the asymmetric
orbifold. Recall that the asymmetric orbifold could be described as a
simple current extension of a tensor product of $\suhat(2)_1$ theories 
\Jdecomp. In the previous section we have shown how to construct the
boundary states for this tensor product theory. Now we need to
implement the simple current extension. Let us begin by defining
an action of $J$ on the set of $\sigma$-twisted representations by  
\eqn\Ssimplecurrent{
\hat{S}^\sigma_{J{\bf \hat{m}},{\bf l}} = 
{J_{\bf l}} \, \hat{S}^\sigma_{{\bf \hat{m}},{\bf l}}\,.
}
Given the structure of $\hat{S}^\sigma$ described in appendix~B, this
prescription defines the action of $J$ uniquely.

The boundary states can now be constructed as for usual simple current
extensions \refs{\FSone,\FStwo}. There are two cases to
distinguish. If a given twisted weight ${\bf \hat{m}}$ is not a fixed
point under the action of $J$, the boundary state is defined by  
\eqn\Bnofixed{
|\!| [{\bf \hat{m}}] \rangle\!\rangle^\sigma = 
{1\over \sqrt{2}} \left(|\!| {\bf \hat{m}} \rangle\!\rangle^\sigma
+ |\!| J{\bf \hat{m}} \rangle\!\rangle^\sigma \right) \,.
}
Here $|\!| {\bf \hat{m}} \rangle\!\rangle^\sigma$ and 
$|\!| J{\bf \hat{m}} \rangle\!\rangle^\sigma$ are boundary states of
the diagonal $\suhat(2)_1^5$ theory defined by \sigmaCardy. The sum in 
\Bnofixed\ guarantees that only $J$-invariant Ishibashi states
contribute. These boundary states therefore only involve Ishibashi
states that come from the first sum in \Jdecomp. They are 
labelled by $J$-orbits, where $[{\bf \hat{m}}]$ denotes the orbit with 
representative ${\bf \hat{m}}$.   

On the other hand, if ${\bf \hat{m}}$ is invariant under the action of
$J$ the above construction has to be modified. A
$\sigma$-twisted representation ${\bf \hat{m}}$ is invariant under
$J$ if and only if $\sigma$ has the property that 
${J_{\bf l}}=+1$ for all ${\bf l}$ in the $\suhat(2)_1^5$ theory
for which $\sigma({\bf l})={\bf l}$. (The simplest example for such a
permutation is $\sigma=(1345)$.) If this is the case, then there 
exist non-trivial Ishibashi states from the $J$-twisted sector, \ie\
from the second sum in \Jdecomp, and conversely, whenever such
Ishibashi states from the $J$-twisted sector exist, the corresponding
permutation has fixed points. In fact, for each such permutation there
is an equal number of Ishibashi states from the $J$-untwisted and
the $J$-twisted sector of \Jdecomp. In order to see this, observe that
there is a $\sigma$-twisted Ishibashi state from the first sum in
\Jdecomp\ for each representation ${\bf l}$ with $J_{\bf l}=+1$ that
satisfies  
\eqn\Juntwist{
\sigma\, (\,{\bf l}\,)\,=\, {\bf l}\,.
}
On the other hand, there is a $\sigma$-twisted Ishibashi state from
the second sum in \Jdecomp\ for each representation ${\bf l}$ with 
$J_{\bf l}=+1$ that satisfies  
\eqn\Jtwist{
\sigma\, (J\,{\bf {l}})\,=\, {\bf {l}}\,.
}
Every solution to \Jtwist\ can be obtained by adding all solutions of 
\Juntwist\ to one fixed solution, ${\bf {l}}_0$, of \Jtwist. 
(As always, addition is understood mod $2$ here.) In particular, their
number is therefore the same. 

We have now assembled all the notation necessary to describe
the boundary states for those permutations that have $J$-fixed
points. These boundary states are labelled by a $\sigma$-twisted
representation ${\bf \hat{m}}$ together with one additional
sign. Explicitly they are given as 
\eqn\Bfixed{
|\!| {\bf \hat{m}},\pm \rangle\!\rangle_{{\bf l}_0}^\sigma 
= {1\over\sqrt{2}} 
\left( \sum_{\bf l} {\hat{S}^\sigma_{{\bf \hat{m}},{\bf l}} \over 
     \sqrt{S_{{\bf 0},{\bf l}}}} \, 
          | {\bf l}  \rangle\!\rangle^\sigma 
\pm \sum_{\bf l} {\hat{S}^\sigma_{{\bf \hat{m}},{\bf l}} \over 
\sqrt{S_{{\bf 0},{\bf l}}}} \, 
        | {\bf l} +{\bf l}_0  \rangle\!\rangle^\sigma\right)\,.
}
The Ishibashi state $| {\bf l}  \rangle\!\rangle^\sigma$ lies in the 
${\bf l}\otimes {\bf l}$ sector of \Jdecomp, while the Ishibashi
state $| {\bf l} +{\bf l}_0  \rangle\!\rangle^\sigma$ lies in 
$({\bf l}+{\bf l}_0)\otimes J({\bf l}+{\bf l}_0)$. Both sums run over
the $\sigma$-invariant representations of $\suhat(2)_1^5$ with 
$J_{\bf l}=+1$. The set of boundary states is independent of the
special solution ${\bf l}_0$ to \Jtwist; in fact, we have 
\eqn\ambig{
|\!| {\bf \hat{m}},\pm \rangle\!\rangle^\sigma_{\bf l_0} = 
|\!| {\bf \hat{m}},\pm (-1)^{{\bf \hat{m}}\cdot ({\bf l_0}-{\bf l_0'})}
              \rangle\!\rangle^\sigma_{\bf l_0'} \,,}
where the inner product of $\sigma$-invariant and $\sigma$-twisted
representations is defined by
\eqn\convention{
\hat{S}^\sigma_{{\bf \hat{m}},{\bf l}+{\bf k}} 
= (-1)^{{\bf \hat{m}}\cdot {\bf k}}
\hat{S}^\sigma_{{\bf \hat{m}},{\bf l}}\,,}
where ${\bf k}$ is a $\sigma$-invariant representation of 
$\suhat(2)_1^5$.

\subsec{The analysis of the overlaps}

In the previous subsection we have described the $\sigma$-twisted
boundary states of the asymmetric orbifold. It should be clear from
the description of the twisted $S$-matrix in appendix~B how to write
down these states explicitly. Given these explicit descriptions, it is
easy to check by hand that the boundary states do indeed satisfy the
Cardy condition.

Now we want to give a more structural argument to this
effect. Since there are two types of boundary states (namely those
defined by \Bnofixed\ and those defined by \Bfixed), there are
three cases to consider. First we analyse the overlaps between two
boundary states of the form \Bnofixed. Using \NIMsutwo\ it follows
that 
\eqn\cardyone{\eqalign{
{}^\sigma\langle\!\langle [{\bf \hat{n}}] |\!| 
q^{{1\over 2}(L_0+\bar L_0 - c/12)} |\!| 
[{\bf \hat{m}}] \rangle\!\rangle^\tau & = 
{1\over 2}\sum_{\bf \hat{p}} \left( 
{\cal N}_{{\bf \hat{p}},{\bf \hat{m}}}^{{\bf \hat{n}}}
+{\cal N}_{{\bf \hat{p}},{\bf \hat{m}}}^{J{\bf \hat{n}}}
+{\cal N}_{{\bf \hat{p}},J{\bf \hat{m}}}^{{\bf \hat{n}}}
+{\cal N}_{{\bf \hat{p}},J{\bf \hat{m}}}^{J{\bf \hat{n}}} \right) \, 
\chi_{\bf \hat{p}}(\tilde{q}) \cr
& = \sum_{\bf \hat{p}} \left(
{\cal N}_{{\bf \hat{p}},{\bf \hat{m}}}^{{\bf \hat{n}}} 
+{\cal N}_{{\bf \hat{p}},J{\bf \hat{m}}}^{{\bf \hat{n}}} \right)\, 
\chi_{\bf \hat{p}}(\tilde{q})\,,}}
where we have dropped the superscripts $(\sigma,\tau)$ and have used
that  
\eqn\Jonfusion{
{\cal N}_{{\bf \hat{p}},J{\bf \hat{m}}}^{J{\bf \hat{n}}}\,
=\, \sum_{\bf l} 
{\left(\hat{S}^\sigma_{J{\bf \hat{n}},{\bf l}} \right)^\ast
 \hat{S}^\tau_{J{\bf \hat{m}},{\bf l}} 
 \hat{S}^{\sigma \tau^{-1}}_{{\bf \hat{p}},{\bf l}} \over
  S_{{\bf 0},{\bf l}}} \,
=\,  \sum_{\bf l} {J_{\bf l}} \left(J_{\bf l}\right)^\ast
{\left(\hat{S}^\sigma_{{\bf \hat{n}},{\bf l}}\right)^\ast 
\hat{S}^\tau_{{\bf \hat{m}},{\bf l}} 
\hat{S}^{\sigma \tau^{-1}}_{{\bf \hat{p}},{\bf l}} \over
S_{{\bf 0},{\bf l}}} \,
=\,{\cal N}_{{\bf \hat{p}},{\bf \hat{m}}}^{{\bf \hat{n}}} \,.
}
The case when one of the two boundary states is of the form \Bfixed,
is very similar. In this case one finds that 
\eqn\cardytwo{\eqalign{
{}^\sigma_{{\bf l}_0}\langle\!\langle {\bf \hat{n}},\pm |\!| 
q^{{1\over 2}(L_0+\bar L_0 - c/12)} |\!| 
[{\bf \hat{m}}] \rangle\!\rangle^\tau & = 
{1\over 2}\sum_{\bf \hat{p}} \left( 
{\cal N}_{{\bf \hat{p}},{\bf \hat{m}}}^{{\bf \hat{n}}}
+{\cal N}_{{\bf \hat{p}},J{\bf \hat{m}}}^{{\bf \hat{n}}} \right) \,
\chi_{\bf \hat{p}}(\tilde{q}) \cr
& = {1\over 2}\sum_{\bf \hat{p}} \left( 
{\cal N}_{{\bf \hat{p}},{\bf \hat{m}}}^{{\bf \hat{n}}}
+{\cal N}_{{\bf \hat{p}},{\bf \hat{m}}}^{J{\bf \hat{n}}} \right) \,
\chi_{\bf \hat{p}}(\tilde{q}) \cr
& = \sum_{\bf \hat{p}} 
{\cal N}_{{\bf \hat{p}},{\bf \hat{m}}}^{{\bf \hat{n}}} 
\, \chi_{\bf \hat{p}}(\tilde{q})\,,}}
where we have used \Jonfusion\ in the second line, and the invariance
of ${\bf \hat{n}}$ in the last. 

Finally, suppose that both boundary states are of the form \Bfixed,
with `in'-state 
$|\!| {\bf \hat{n}},\pm \rangle\!\rangle^\sigma_{{\bf l}_0}$ 
and `out'-state 
$|\!| {\bf \hat{m}},\pm \rangle\!\rangle^\tau_{{\bf k}_0}$.  By 
the same argument as in \cardytwo, the contribution from the first sum 
in \Bfixed\ gives rise to 
\eqn\cardythreo{
{1\over 2} \sum_{\bf {\hat{p}}} 
{\cal N}_{{\bf \hat{p}},{\bf \hat{m}}}^{{\bf \hat{n}}} 
\, \chi_{\bf \hat{p}}(\tilde{q})\,.}
On the other hand, the contribution from the second sum in \Bfixed\
depends on whether there are weights ${\bf l}$ in $\suhat(2)_1^5$
that satisfy simultaneously \Jtwist\ with $\sigma$ and $\tau$. If such
a weight ${\bf l}_0'$ exists, then the contribution from the second sum
in \Bfixed\ gives 
\eqn\cardythret{\eqalign{
(-1)^{{\bf \hat{m}}\cdot ({\bf l}_0 - {\bf l}_0')}
{1\over 2} & \sum_{\bf l} \sum_{\bf \hat{p}}
{\left(\hat{S}^\sigma_{{\bf \hat{n}},{\bf l}}\right)^\ast \,
\hat{S}^\tau_{{\bf \hat{m}},{\bf l}+{\bf l}_0'-{\bf k}_0} \,
 \hat{S}^{\sigma \tau^{-1}}_{{\bf \hat{p}},{\bf l}+{\bf l}_0'} \over
  S_{{\bf 0},{\bf l}}} \, \chi_{{\bf \hat{p}}}(\tilde{q}) \cr
& = {1\over 2} 
(-1)^{{\bf \hat{m}}\cdot ({\bf l}_0-{\bf k}_0)}
\sum_{\bf \hat{p}} (-1)^{{\bf \hat{p}} \cdot {\bf l}_0'}
\sum_{\bf l} 
{\left(\hat{S}^\sigma_{{\bf \hat{n}},{\bf l}}\right)^\ast \,
\hat{S}^\tau_{{\bf \hat{m}},{\bf l}} \,
 \hat{S}^{\sigma \tau^{-1}}_{{\bf \hat{p}},{\bf l}} \over
  S_{{\bf 0},{\bf l}}} \, \chi_{{\bf \hat{p}}}(\tilde{q}) \cr
& = {1\over 2} 
(-1)^{{\bf \hat{m}}\cdot ({\bf l}_0-{\bf k}_0)}\,
\sum_{\bf \hat{p}} (-1)^{ {\bf \hat{p}} \cdot {\bf l}_0'} \,
{\cal N}_{{\bf \hat{p}},{\bf \hat{m}}}^{{\bf \hat{n}}}
\, \chi_{{\bf \hat{p}}}(\tilde{q}) \,,
}}
where we have used \ambig. Depending on whether the signs of the two
boundary states are the same or opposite, \cardythret\ has to be added
or subtracted from \cardythreo. In either case, taking both terms
together we obtain a non-negative integer linear combination of
characters in the open string.

Finally, if none of the $\suhat(2)_1^5$ weights satisfies \Jtwist\
simultaneously for $\sigma$ and $\tau$, the contribution from the
second sum in \Bfixed\ vanishes. If this is the case, then one can
show that for one of the sets ${\cal F}_i$ defined in appendix~B,
$m$ in (B.6) is non-zero. This implies that the coefficients 
\cardythreo\ are actually even integers. The coefficients in
\cardythreo\ are then again integers, thus proving the Cardy
condition. 

\newsec{Some generalisations}

Up to now we have only discussed the D-branes that preserve the
$\suhat(2)_1^5$ symmetry up to outer automorphisms, \ie\ that
satisfy \generalglue\ with $\sigma\in S_5$. It is relatively
straightforward to include also inner automorphisms. The group of
inner automorphisms is isomorphic to $SU(2)^5$, and it acts on the
algebra $\suhat(2)_1^5$ by conjugation. The induced action on the
boundary states is simply given by the global action of $SU(2)^5$ on
the right-moving states, say. Since this action corresponds to a
marginal deformation by a local field, the resulting boundary states
satisfy the relevant consistency conditions \refs{\RS}.  

It was shown in \refs{\GRW} that the D-branes that preserve the
conformal symmetry $\Vir$ for $\suhat(2)_1$, necessarily
preserve the full chiral algebra  $\suhat(2)_1$ up to an inner
automorphism. It therefore follows that the D-branes that preserve
$\suhat(2)_1^5$ up to an inner automorphism in $SU(2)^5$ account
already for all D-branes that preserve $\Vir^5$. We have therefore
managed to construct all D-branes that preserve $\Vir^5$ up to 
the outer automorphisms that are isomorphic to $S_5$.
\medskip

The orbifold we have been considering acts as (some version of)
T-duality on the four-torus part. One would therefore expect that the
theory has D-branes that are simply superpositions of Dp-D(4-p)
branes. In addition to this T-duality, there is the shift action 
along the additional circle; the pairs of branes will therefore be
localised at opposite points along the fifth circle. 

The combinations of Dp-D(4-p) branes preserve the full chiral algebra
\diagonal\ up to certain inner automorphisms. In terms of the original 
description of the orbifold theory \HUHT, the corresponding boundary
states therefore only involve Ishibashi states from the untwisted
sector. Following the discussion leading to \Jdecomp, they correspond
to $(25)$-twisted boundary states in the second formulation of the
theory. Since $\sigma=(25)$ does not have any fixed points (in the
sense discussed above in section~4), the corresponding boundary states
then also only involve Ishibashi states from the first sum in
\Jdecomp. Of the $32=2^5$ representations of 
$\suhat(2)_1^5$, $16=2^4$ representations are invariant under the
action of $(25)$. There are therefore sixteen $(25)$-twisted
representations of $\suhat(2)_1^5$, each of which lies in an orbit of
length two under the action of $J$. Thus there are eight different
boundary states that satisfy \generalglue\ with $\sigma=id$. We want
to show that these eight boundary states can be identified with
combinations of D0-D4 branes, where the position of the D4-brane is
shifted along the fifth circle relative to the position of the
D0-brane. 

Recall from appendix~B that the $(25)$-twisted $S$-matrix is simply a
product of four $S$-matrices of $\suhat(2)_1$. One of the
$(25)$-twisted representations, that we shall denote by $\hat{\bf 0}$
in the following, has thus the property that 
\eqn\nulldef{
\hat{S}^{(25)}_{\hat{\bf 0},{\bf l}} = {1\over \sqrt{2}^4} = 
{1\over 4}\,,}
for all $(25)$-invariant representations ${\bf l}$ of
$\suhat(2)_1^5$. The boundary state associated to $\hat{\bf 0}$ via  
\Bnofixed, $|\!| [{\bf \hat{0}}] \rangle\!\rangle^{(25)}$, is
therefore simply the sum (with overall normalisation $2^{-1/4}$ but
without signs) of the eight $(25)$-twisted Ishibashi states coming
from the first sum of \Jdecomp. Alternatively, it is the same sum over
the eight  (untwisted) Ishibashi states coming from the untwisted
sector of \HUHT.  

The same boundary state can now be obtained starting from the original
$\sohat(8)_1\oplus\suhat(2)_1$ theory as follows. Consider the Cardy
boundary state associated to the representation $(o,{\bf 0})$ of 
$\sohat(8)_1\oplus\suhat(2)_1$. From a geometrical point of view, this
boundary state describes a single D0-brane. By the usual Cardy formula
the boundary state is the sum (with overall normalisation $2^{-3/4}$
but without signs) over all eight Ishibashi states of
$\sohat(8)_1\oplus\suhat(2)_1$. Because of the decomposition 
\soeightintosofour, each such Ishibashi state is the sum of two 
$\suhat(2)_1^5$ Ishibashi states, and therefore the Cardy state
corresponding to $(o,{\bf 0})$ is the sum of sixteen Ishibashi states of
$\sohat(8)_1\oplus\suhat(2)_1$ (with overall normalisation $2^{-3/4}$
but without signs). Finally, when we impose the orbifold projection,
only half of these Ishibashi states survive, and the overall
normalisation becomes $\sqrt{2}\, 2^{-3/4} = 2^{-1/4}$. The resulting
boundary state therefore agrees with the boundary state 
$|\!| [{\bf \hat{0}}] \rangle\!\rangle^{(25)}$ described above. On the
other hand, the action of the orbifold acts geometrically on the
$SO(8)$ lattice and the circle, and simply maps the D0-brane to a
D4-brane located at the opposite point of the extra circle. 

We have therefore shown that the boundary state 
$|\!| [{\bf \hat{0}}] \rangle\!\rangle^{(25)}$ describes the boundary
state 
\eqn\interp{
|\!| [{\bf \hat{0}}] \rangle\!\rangle^{(25)} = {1\over\sqrt{2}} 
\left( |\!| \hbox{D0},x \rangle\!\rangle + 
       |\!| \hbox{D4},x+\pi R_5 \rangle\!\rangle \right) \,.}
Similarly, the other seven $(25)$-twisted boundary states can be
obtained by the same construction starting with the Cardy state
corresponding to some other representation of
$\sohat(8)_1\oplus\suhat(2)_1$; they therefore describe combinations
of D0-D4 branes where the position of the D0-brane is at a different
point in the $SO(8)$ torus. It should also be clear that the
combinations of Dp-D(4-p) branes can be obtained from the above by the
action of the inner automorphism of $SU(2)^5$.

\newsec{Conclusions}

In this paper we have determined the boundary states for an
asymmetric orbifold of the bosonic string. More precisely, we have
constructed all the boundary states that preserve five copies of the
Virasoro algebra at $c=1$ up to permutations. The corresponding
D-branes include the usual superpositions of Dp-D(4-p) branes that
only couple to the untwisted sector of the asymmetric
orbifold. However, we have also constructed branes that couple to
twisted sector states. 

The boundary states that only couple to the untwisted sector of the
orbifold involve at most eight Ishibashi states; their overall
normalisation (which is proportional to the tension) is therefore at
least $2^{-1/4}$. These boundary states therefore do not describe the
`lightest' D-branes. Indeed, some of the boundary states (for example
those that correspond to $\sigma=(25)$ in the first formulation of the
theory) involve sixteen Ishibashi states, and the overall
normalisation is then $2^{-3/4}$. It would be interesting to
understand the geometrical interpretation of these boundary states. 

The techniques we have employed in our construction should generalise
to other asymmetric orbifolds. In particular, it would be interesting
to apply these ideas to superstring orbifolds, for example the one
considered in  \refs{\Harvey}.

\vskip 1cm

\centerline{{\bf Acknowledgments}}
\pano
We thank Ilka Brunner, Peter Goddard, Axel Kleinschmidt, Christian
Stahn and in particular Andreas Recknagel for useful discussions. 

MRG is grateful to the Royal Society for a University Research
Fellowship. He thanks the Institute for Advanced Study for hospitality
while this paper was being completed; his research there was supported
by a grant in aid from the Funds for Natural Sciences. S.S.-N. is
grateful to St.\ John's College, Cambridge, for a Jenkins
Scholarship. We also acknowledge  partial support from the PPARC
Special Programme Grant `String Theory and Realistic Field Theory',
PPA/G/S/1998/0061 and the EEC contract HPRN-2000-00122.

\vskip1cm


\appendix{A}{Character and Theta-function conventions}

In this paper we are using the theta functions
\eqn\Thetas{
\Theta_{m,k}(q)=\sum_{n\in\Zop+ {m\over 2k}}q^{n^2 k}\,,}
which have the modular transformation properties
\eqn\modular{
S \Theta_{m,k}= \left({i\tau\over 2k} \right)^{\half}
\sum_{n\in\Zop / 2k} e^{-inm\pi/k}\,  \Theta_{n,k}
\qquad \hbox{ and }\qquad
T \Theta_{m,k}= e^{i\pi m^2 /2k} \, \Theta_{m,k}\,.}
The two characters of $\suhat(2)_1$ corresponding to $j=0$ and $j=1/2$
are denoted by $\chi_{0}$ and $\chi_1$, respectively. They are
related to the theta-functions \Thetas\ as
\eqn\chiinThetas{
\chi_0(q)\,=\, {\Theta_{0,1}(q)\over \eta(q)}\,,\qquad 
\chi_1(q)\,=\, {\Theta_{1,1}(q)\over\eta(q)}\,,} 
where $\eta(q)$ is the Dedekind eta-function
\eqn\ded{
\eta(q)= q^{-{1\over 24}} \prod_{n=1}^{\infty} (1-q^n) \,.}
The modular $S$-matrix of $\suhat(2)_1$ is then given by 
\eqn\ssutwo{
S_{\suhat(2)_1}\,=\,{1\over\sqrt{2}}
\left(\matrix{1 & 1 \cr 1& -1}\right)\,.
}
The functions \Thetas\ are related to the usual $\theta_i$
Jacobi theta-functions as
\eqn\ThetastoJacobies{\eqalign{
\theta_{2}(\tau)\,&
=\,\Theta_{1,1}(\tau/2)\,=\,\sqrt{\Theta_{0,1}(\tau)\Theta_{1,1}(\tau)}\,=\, 
\Theta_{1,2}(\tau)+ \Theta_{3,2}(\tau) \cr 
\theta_{3}(\tau)\,& =\, \Theta_{0,1}(\tau/2)\, =\,
\Theta_{0,4}(\tau/2)+\Theta_{4,4}(\tau/2)\,=\,
\sqrt{\Theta_{0,1}^2(\tau) + \Theta_{1,1}^2(\tau) }\cr
&= \Theta_{0,2}(\tau)+ \Theta_{2,2}(\tau) \cr
\theta_{4}(\tau)\,& =\, \Theta_{0,1}(\tau/2, z=1/2)\,=\,
\Theta_{0,4}(\tau/2)-\Theta_{4,4}(\tau/2)\,=\,
\sqrt{\Theta_{0,1}^2(\tau) - \Theta_{1,1}^2(\tau) }\cr
&= \Theta_{0,2}(\tau)-\Theta_{2,2}(\tau) \,.
}}
In terms of these, the (specialised) characters for $\sohat(2p)_1$, 
$p\in \Nop$, are 
\eqn\socharacters{
O_{2p}\, =\, {\theta_3^{p} + \theta_4^{p}\over 2 \eta^{p} }\,,\quad
V_{2p}\, =\, {\theta_3^{p} - \theta_4^{p}\over 2 \eta^{p} }\,,\quad
S_{2p}\, =\, C_{2p}\,=\,  {\theta_2^{p}\over 2 \eta^{p} }\,,}
and their modular transformation matrices are 
\eqn\soS{
S_{so(2p)}\, =\, \half\, \left(
\matrix{
1&1&1&1\cr
1&1&-1&-1\cr
1&-1& e^{-ip\pi/2} & -e^{-ip\pi/2}\cr
1&-1& -e^{-ip\pi/2} & e^{-ip\pi/2}
}\right) \,,\quad 
T_{so(2p)}\, =\,\left(
\matrix{
1&0&0&0\cr
0&-1&0&0\cr
0&0& e^{ip\pi/4} & 0\cr
0&0&0& e^{ip\pi/4}}\right)\,.}


\appendix{B}{The Cardy condition for the permutation-twisted boundary
states} 

Let us first compute the $\sigma$-twisted $S$-matrix for
$\suhat(2)_1^n$, where $\sigma\in S_n$. Suppose first that
$\sigma$ consists of a single non-trivial cycle of length $k>1$. Then
there are $2^{n-k+1}$ $\sigma$-invariant representations of
$\suhat(2)_1^n$; they can be labelled by $(l_1,\ldots,l_{n-k},l)$,
where $l$ is the representation label for the $k$ representations that
are permuted among each other by $\sigma$. The twining character is
then  
\eqn\twinex{
\chi_{(l_1,\ldots,l_{n-k},l)}^{(\sigma)}(q) = 
\chi_l(q^k) \, \prod_{i=1}^{n-k} \chi_{l_i}(q) \,.}
Upon a modular transformation this becomes 
\eqn\twismod{
\chi_{(l_1,\ldots,l_{n-k},l)}^{(\sigma)}(q) = 
\sum_{(m_1,\ldots,m_{n-k},\hat{m})} 
S_{l,\hat{m}} \,\chi_{\hat{m}}(\tilde{q}^{1/k}) \, 
\prod_{i=1}^{n-k} S_{l_i,m_i} \chi_{m_i}(\tilde{q}) \,,}
where all $S$-matrices here are the $S$-matrix of $\suhat(2)_1$. 
The characters $\chi_{\hat{m}}(\tilde{q}^{1/k})$ can be identified
with the $\sigma$-twisted characters of $\suhat(2)_1^k$, and thus the
twisted $S$-matrix is simply a suitable product of $S$-matrices of 
$\suhat(2)_1$. It is easy to see that this argument generalises
directly to the case of an arbitrary permutation: if $\sigma$ has
$c_\sigma$ cycles (counting trivial cycles) then the
$\sigma$-invariant representations are labelled by $c_\sigma$ labels
(each of which can take the values $0,1$), and the $\sigma$-twisted
$S$-matrix is the product of $c_\sigma$ $S$-matrices of
$\suhat(2)_1$.

With these preparations we can now show that the coefficients
\twistedfusion  
\eqn\gtf{
{\cal N}_{{\bf \hat{p}},{\bf \hat{m}}}^{{\bf \hat{n}}}= 
\sum_{\bf l}
{\left(\hat{S}^\sigma_{{\bf \hat{n}},{\bf l}}\right)^\ast 
\hat{S}^\tau_{{\bf \hat{m}},{\bf l}} 
\hat{S}^{\sigma \tau^{-1}}_{{\bf \hat{p}},{\bf l}} \over
S_{{\bf 0},{\bf l}}}
}
define indeed non-negative integers. (The following argument is
similar to the argument given in appendix~A of \recknagel.) The idea
of the argument is to reduce this expression to a product of
conventional fusion rules of $\suhat(2)_1$, using the Verlinde
formula. The main problem in doing so is that the sum over ${\bf l}$
only runs over those indices ${\bf l}=(l_1,\ldots,l_n)$ that are
simultaneously invariant under $\sigma$ and $\tau$. Furthermore, if we
write out the twisted $S$-matrices in the numerator in terms of the
$S$-matrices of $\suhat(2)_1$, there are only
$c_\sigma+c_\tau+c_{\sigma \tau^{-1}}$ $S$-matrices, whereas we would
need $3n$ $S$-matrices in order to group them into Verlinde-formula
expressions.  

For each $i\in\{1,\ldots,n\}$ let us denote by ${\cal F}_i$ the subset
of labels in $\{1,\ldots,n\}$ that have to take the same value as
$l_i$ in the sum over ${\bf l}$ above, \ie\ $l_j=l_i$ for 
$j\in{\cal F}_i$. In order to prove the above formula we can consider
each such set at a time (since the total formula will just be the
product of the expressions corresponding to each such set). Without
loss of generality, let $i=1$, and let ${\cal F}_1=\{1,\ldots,r\}$. 
Let us restrict $\sigma$, $\tau$ and $\sigma \tau^{-1}$ to the set 
${\cal F}_1=\{1,\ldots,r\}$, and let $s$ be the number of cycles of
$\sigma$ among ${\cal F}_1$ (including trivial cycles). Similarly,
define $t$ to be the number of cycles of $\tau$, and $u$ the number of
cycles of $\sigma \tau^{-1}$. The contribution to \gtf\
coming from ${\cal F}_1$ is then  
\eqn\partialcont{{\cal N}' = 
\sum_{l=0,1} {S_{n_1,l} \cdots S_{n_s,l} \, 
S_{m_1,l} \cdots S_{m_t,l} \,
S_{p_1,l} \cdots S_{p_u,l} \over
S_{0,l} \cdots S_{0,l}}\,,}
where the product in the denominator contains $r$ powers of $S_{0,l}$,
and we have used that $S$ is real. Next we want to turn this into $r$
sums by inserting the identity
\eqn\Sindet{
\sum_a S_{a,l} S_{a,l'} = \delta_{l,l'}}
$r-1+m$ times, where $m$ is the non-negative integer defined by 
\eqn\group{
s+t+u+2m = r+2 \,.}
(As we shall see momentarily, \group\ defines indeed a non-negative 
integer $m$.) We can then distribute the remaining factors of $S$ as   
\eqn\partialcontd{
{\cal N}' = 
\sum_{l_1,\ldots,l_{r}} \sum_{a_1,\ldots,a_{r-1+m}}
{S_{n_1,l_1} S_{m_1,l_1} S_{a_1,l_1} \over S_{0,l_1}}
{S_{a_1,l_2} S_{n_2,l_2} S_{a_2,l_2} \over S_{0,l_2}} \cdots 
{S_{a_{r-1+m},l_{r}} S_{m_t,l_{r}} S_{p_u,l_{r}} \over S_{0,l_{r}}} \,.}
Each sum over $l_i$ gives now a fusion rule coefficient via the
Verlinde formula, and it is therefore manifest that ${\cal N}'$ is a
non-negative integer. 

It therefore only remains to prove \group. This can be done by
induction on $r$. The case $r=1$ is trivial. Assume therefore that the
statement holds for $r-1$, and let $\sigma$ and $\tau$ be as
above. Then we can find transpositions $(jr)$ and $(kr)$ (where either
$j$ or $k$ but not both may be equal to $r$) so that
$\sigma=(jr) \sigma'$ and $\tau=(kr)\tau'$, where $\sigma'$ and $\tau'$
leave $r$ invariant. Thus $\sigma'$ and $\tau'$ satisfy the
assumptions of the statement with $r-1$, and we have that 
$s'+t'+u' + 2m'=r+1$, where $s'$, $t'$, and $u'$ are defined in the obvious 
manner. By construction 
\eqn\stprime{
s' = \left\{
\eqalign{ s \quad & \quad  \hbox{if $j\ne r$} \cr
          s-1 & \quad \hbox{if $j=r$}} \right. \qquad
t' = \left\{
\eqalign{ t \quad & \quad  \hbox{if $k\ne r$} \cr
          t-1 & \quad  \hbox{if $k=r$.}}\right. }
The number of cycles of a permutation is the same in each conjugacy
class, and therefore $u$ is equal to the number of cycles of the
permutation 
\eqn\ucycle{
(kr) \sigma \tau^{-1} (kr) = (kr) (jr) \sigma' \tau'^{-1} \,.}
Now the product $(kr)(jr)$ is equal to 
\eqn\productcycle{
(kr)(jr) = \left\{ \eqalign{
(kr) & \quad \hbox{if $j=r$} \cr
(jr) & \quad \hbox{if $k=r$} \cr
\hbox{id} & \quad \hbox{if $j=k\ne r$} \cr 
(jkr) & \quad \hbox{otherwise.} } \right. }
Thus it follows from \ucycle\ that 
\eqn\uprime{
u' = \left\{ 
\eqalign{ u \quad & \quad \hbox{if $j=r$ or $k=r$} \cr
u-1 & \quad \hbox{if $j=k\ne r.$}}\right.} 
So if $j=r$ or $k=r$ or $j=k\ne r$, then $s'+t'+u'=s+t+u-1$, and the
induction step follows (with $m=m'$). This leaves us with analysing
the case when $j\ne k$ with neither $j$ nor $k$ equal to $r$. The
answer then depends on whether $j$ and $k$ lie in the same cycle of 
$\sigma'\tau'^{-1}$ or whether they do not. In the former case, the
permutation in \ucycle\ is 
\eqn\ucycletwo{\eqalign{
(jkr) \, & (j\, v_1 v_2 \cdots v_L \, k\, w_1 w_2 \cdots w_M) \,
\hbox{(other cycles)} \cr
& = (j\, v_1 v_2 \cdots v_L r) \, (k\, w_1 w_2 \cdots w_M) \,
\hbox{(other cycles)}\,,}}
and thus $u'=u-1$, and hence $s'+t'+u' = s+t+u -1$. As before the 
induction step then follows with $m=m'$. In the other case we have
instead of \ucycletwo 
\eqn\ucycleone{\eqalign{
(jkr) \, &  (j\, v_1 v_2 \cdots v_L) \,  (k \, w_1 w_2 \cdots w_M) \,
\hbox{(other cycles)} \cr
& = (j \, v_1 v_2 \cdots v_L \,k \,w_1 w_2 \cdots w_M r) \,
\hbox{(other cycles)}\,,}}
and thus $u' = u+1$, and hence $s'+t'+u' = s+t+u +1$. In this case the
induction step follows with $m=m'+1$. This proves the statement \group.

\appendix{C}{Twisted fusion rules}

In this appendix we want to demonstrate that for a simple example 
\twistedfusion\ does indeed describe the twisted fusion rules. The
example we want to consider is $\suhat(2)_1^2$, where $\sigma$ is the
transposition that exchanges the two factors of $\suhat(2)_1$. As was
explained in the previous appendix, there are two $\sigma$-twisted
representations. Every $\sigma$-twisted representation of
$\suhat(2)_1^2$ defines an untwisted representation of
$\suhat(2)_2\oplus \Vir_{1/2}$, where $\Vir_{1/2}$ is the Virasoro
algebra at $c=1/2$.\footnote{$^\dagger$}{In the general case where
$\sigma$ is a permutation acting on $\suhat(2)_1^n$ with $c$ cycles
of length $l_i\geq 1$, $i=1,\ldots, c$, the relevant chiral algebra is 
$$
\bigoplus_{i=1}^{c} \suhat(2)_{l_i} \oplus {\cal M}_{rem}\,,
$$
where ${\cal M}_{rem}$ is a chiral algebra of suitable central
charge.} 
In general, an irreducible $\sigma$-twisted
representation of $\suhat(2)_1^2$ will contain a number of irreducible
representations of $\suhat(2)_2\oplus \Vir_{1/2}$.  

In order to determine the characters of the twisted representations
let us consider the twining characters 
\eqn\leveltwocharacters{\eqalign{
\hbox{Tr}_{\H_0\otimes \H_0}(\sigma q^{L_0-{c\over 24}}) & = 
\chi_{0}(q^2)= {\Theta_{0,2}(q)\over \eta(q^2)} \, =\, 
\chi_0^{\,(2)}(q)\, \chi_{0}(q)\,
-\, \chi_2^{\,(2)}(q)\, \chi_{1/2}(q)\,,\cr
\hbox{Tr}_{\H_1\otimes \H_1}(\sigma q^{L_0-{c\over 24}}) & =
\chi_{1}(q^2)\,=\, {\Theta_{2,2}(q)\over \eta(q^2)} \,=\,- 
\chi_0^{\,(2)}(q)\, \chi_{1/2}(q)\,
+\, \chi_2^{\,(2)}(q)\, \chi_{0}(q)\,,}}
which we have written in terms of (conventional) characters of  
$\suhat(2)_2\, \oplus\, \Vir_{1/2}$. (The characters of $\suhat(2)_2$
are denoted by $\chi_j^{\,(2)}(q)$, while the characters of
$\Vir_{1/2}$ are $\chi_{h}(q)$.) Upon an S-modular transformation we
then find 
\eqn\Sonleveltwocharacters{\eqalign{
S\, \chi_{\H_0\otimes \H_0}^{(\sigma)}\,
=& {(\theta_3^{3/2}\theta_2^{1/2} 
+ \theta_3^{1/2}\theta_2^{3/2})\over 2\eta^2}\cr 
= & \, {1\over\sqrt{2}}\, \left(\, 
(\chi_0^{\,(2)}+ \chi_2^{\,(2)})
\chi_{1/16}
+ \chi_1^{\,(2)}(\chi_{0}+\chi_{1/2})\,\right)\,,\cr 
S\, \chi_{\H_1\otimes \H_1}^{(\sigma)}\,
=& {(\theta_3^{3/2}\theta_2^{1/2} -\theta_3^{1/2}\theta_2^{3/2} )\over
2\eta^2}\cr 
=&\, {1\over\sqrt{2}}\, \left(\, (\chi_0^{\,(2)}+
\chi_2^{\,(2)})\chi_{1/16}-
\chi_1^{\,(2)}(\chi_{0}+\chi_{1/2})\,\right) \,.
}}
The characters of the twisted representations are therefore 
$\chi_U=\chi_1^{\,(2)}(\chi_{0}+\chi_{1/2})$ 
and $\chi_V=(\chi_0^{\,(2)}+ \chi_2^{\,(2)})\chi_{1/16}$, and 
the $\hat{S}$-matrix is given by 
\eqn\shatmatrixleveltwo{
\hat{S}\,=\,{1\over\sqrt{2}}
\left(\matrix{1 & 1 \cr 1& -1}\right)\,,
}
in agreement with the discussion of the previous
appendix. Furthermore, the two twisted representations $U$ and $V$
decompose with respect to $\suhat(2)_2\oplus \Vir_{1/2}$ as 
\eqn\twisdec{
U = \H_1^{(2)} \otimes \left( \H_{0} \oplus \H_{1/2} \right)
\,,\qquad 
V = \left(\H_0^{(2)} \oplus \H_2^{(2)}\right) \otimes \H_{1/16}\,.
}

The conjectured formula for the twisted fusion rules \twistedfusion\
now predicts that the fusion rules are 
\eqn\twistedfusionp{
\N_{(0,0)} = \N_{(1,1)} = \pmatrix{1 & 0 \cr 0 & 1} \,, \qquad
\N_{(1,0)} = \N_{(0,1)} = \pmatrix{0 & 1 \cr 1 & 0} \,,}
where $(l,m)$ labels the untwisted representation of $\suhat(2)_1^2$,
and the matrix acts on the space of twisted representations with basis
$U$ and $V$. These fusion rules are in agreement (and could have been
derived) from the description of these representations in terms of 
$\suhat(2)_2\, \oplus\, \Vir_{1/2}$. For example, the representation
$(1,1)$ corresponds to $(j=0,h=1/2)\oplus(j=2,h=0)$ of 
$\suhat(2)_2\,\oplus\,\Vir_{1/2}$, and therefore the fusion of
$(1,1)$ with $U(V)$ can only contain $U(V)$. Similarly, $(1,0)$
corresponds to $\left(j=1,h={1\over 16}\right)$ of 
$\suhat(2)_2\,\oplus\,\Vir_{1/2}$, and the fusion of $(1,0)$ with
$U(V)$ can therefore only contain $V(U)$. The other two cases are
identical.


\listrefs

\bye